%% file: main.tex
\documentclass[a4paper,onecolumn,11pt,accepted=2024-03-16]{quantumarticle}
\pdfoutput=1

\include{structure}

\Crefname{equation}{Eq.}{}
\Crefname{figure}{Fig.}{}
\title{Overhead-constrained circuit knitting for variational quantum dynamics}

\author{Gian Gentinetta}
\orcid{0000-0002-5891-3289}
\email{gian.gentinetta@epfl.ch}
\author{Friederike Metz}
\orcid{0000-0002-4745-7329}
\author{Giuseppe Carleo}
\orcid{0000-0002-8887-4356}
\affiliation{Institute of Physics, \'Ecole Polytechnique F\'ed\'erale de Lausanne (EPFL), CH-1015 Lausanne, Switzerland \\
Center for Quantum Science and Engineering, \'Ecole Polytechnique F\'ed\'erale de Lausanne (EPFL), CH-1015 Lausanne, Switzerland}

 \date{}

\begin{document}
\maketitle

\begin{abstract}
Simulating the dynamics of large quantum systems is a formidable yet vital pursuit for obtaining a deeper understanding of quantum mechanical phenomena. While quantum computers hold great promise for speeding up such simulations, their practical application remains hindered by limited scale and pervasive noise. In this work, we propose an approach that addresses these challenges by employing circuit knitting to partition a large quantum system into smaller subsystems that can each be simulated on a separate device. The evolution of the system is governed by the projected variational quantum dynamics (PVQD) algorithm, supplemented with constraints on the parameters of the variational quantum circuit, ensuring that the sampling overhead imposed by the circuit knitting scheme remains controllable. We test our method on quantum spin systems with multiple weakly entangled blocks each consisting of strongly correlated spins, where we are able to accurately simulate the dynamics while keeping the sampling overhead manageable. Further, we show that the same method can be used to reduce the circuit depth by cutting long-ranged gates.
\end{abstract}

\maketitle

\section{Introduction}
Quantum computers are promising tools for simulating quantum systems~\cite{Feynman1982,ibm_17, chiesa_2019, google_2020, arute2020observation, neill2021}. Particularly, the efficient simulation of quantum dynamics can provide insightful information about the nature of physical phenomena at the microscopic scale~\cite{zhang2017,dborin2022,ebadi2021,altman2018,Jong2022,utility2023}. However, the practical utility of quantum devices is currently constrained by limitations in scale and the effects of noise~\cite{childs2018,Babbush2018,Nam2019,Motta2021}. While the size of available quantum computers is steadily growing~\cite{ibm_roadmap}, most publicly available devices are still very limited in size. In order to extend the capabilities of Noisy Intermediate-Scale Quantum (NISQ) devices~\cite{Preskill2018quantumcomputingin}, 
several schemes have been proposed to partition large systems into small clusters that can be solved individually on smaller quantum hardware~\cite{Bravyi_2016, Peng2020, Mitarai_2021,Mitarai2021overheadsimulating, piveteau2023circuit,Fan2015, Gunst2017,yamazaki2018practical, Rossmannek_2021,Eddins_2022, Huembeli2022, paulin2023,gomez2023nearterm, barison2023embedding,XiaoHybridTensor2021,SunPerturbative2022}. To combine the solutions and recover the entanglement between the subsystems, classical resources are usually employed. Hence, ultimately, these hybrid quantum-classical computing approaches allow for quantum simulations on a larger scale.

Developing strategies for efficiently partitioning quantum computations is especially timely, as one of the focuses of the next generation of quantum processors lies in connecting multiple medium-size quantum chips, allowing for parallelization of quantum simulations with real-time classical communication~\cite{ibm_roadmap}. This strategy is of particular utility if each subsystem that is simulated on a separate device is itself highly entangled and, hence, difficult to simulate classically. On the other hand, the entanglement between the partitions should be weak such that classical methods can be efficiently employed to recombine the subsystems. The idea of splitting a quantum system into subsystems can also be motivated by the underlying physical or chemical processes. Several interesting physical systems naturally allow for partitioning into weakly-entangled subsystems such as ground and low-energy eigenstates of local lattice Hamiltonians~\cite{Eisert2010area,Schollw_ck_2011, liu2019vqe} and molecules~\cite{McArdle2020}, as well as quantum impurities immersed in a bath~\cite{Kotliar_2006, Sun2016}.

Two prominent hybrid quantum-classical schemes that combine multiple quantum circuits using classical post-processing are entanglement forging~\cite{Eddins_2022, Huembeli2022, paulin2023} and circuit knitting~\cite{Bravyi_2016, Peng2020, Mitarai_2021,Mitarai2021overheadsimulating, piveteau2023circuit}. Entanglement forging relies on the fact that a bipartite quantum state can always be written in the Schmidt decomposition. This enables a classical computer to combine the states of two systems implemented on separate quantum devices. If the two systems are weakly entangled with each other, a small number of Schmidt coefficients suffices for a good approximation of the full solution.
Crucially, entanglement forging is limited to two subsystems, as the Schmidt decomposition cannot be applied to general multipartite states. 
Circuit knitting, on the other hand, employs quasi-probability distributions to cut gates that span across different systems into locally realizable quantum channels. This allows to arbitrarily cut a quantum circuit into multiple subsystems. However, this technique imposes a sampling overhead that scales exponentially in the number of gates cut.

In this work, we propose a method for quantum time evolution that splits a quantum circuit ansatz into multiple subsystems using circuit knitting while keeping the sampling overhead controlled. This is achieved by imposing a constraint on the circuit parameters during the optimization of the variational quantum circuit.

We employ this method to simulate the dynamics of quantum systems using the projected variational quantum dynamics (PVQD) algorithm~\cite{Barison_2021}. 
While there have been implementations of quantum-classical hybrid schemes to quantum dynamics using perturbation theory~\cite{SunPerturbative2022} or by leveraging mean-field corrections and auxiliary qubits~\cite{gomez2023nearterm}, an application to variational quantum dynamics is largely missing in the literature. 
The task is non-trivial, as evolving a parameterized quantum state in time either requires measuring (complex) matrix elements of the geometric tensor~\cite{dirac_1930, frenkel_1934, mclachlan_1964, Yuan2019theoryofvariational} or fidelities between quantum states~\cite{Barison_2021,gacon2023variational}. 
This poses a challenge to entanglement forging, where the ansatz is given by a superposition of quantum circuits. There, measuring overlaps is expensive and usually requires non-local circuits such as Hadamard-tests~\cite{Cleve_1998}. Instead, in the framework of circuit knitting, fidelities can be straightforwardly computed using, for example, the compute-uncompute method~\cite{Havlicek_2019} without introducing any ancilla qubits or long-ranged gates.

We test our method on spin systems in a transverse field Ising model, where we weakly couple multiple blocks of strongly correlated spins. We show that with a realistic sampling overhead, we can significantly improve the accuracy of the simulation compared to a pure block product approximation, which does not consider any entanglement between different blocks. Furthermore, the trade-off between the sampling overhead and the accuracy of the variational state can be tuned in a controlled way via a single hyperparameter of the optimization. Finally, we demonstrate that our scheme can also reduce the required circuit depth when simulating models containing long-range interactions.

The structure of this paper is as follows: In~\Cref{sec:methods}, we explain how we use PVQD and circuit knitting techniques to evolve a quantum circuit ansatz in time while keeping the sampling overhead controlled. In~\Cref{sec:results}, we test our method on quantum spin systems in a transverse field Ising model for different setups. Finally, in~\Cref{sec:conclusion}, we discuss the results and provide an outlook on possible future applications of the method.

\section{Methods}
\label{sec:methods}
We consider the dynamics of a quantum system represented by a Hilbert space partitioned into $N$ individual subsystems (called blocks) $\mathcal{H} = \mathcal{H}_1 \otimes \mathcal{H}_2 \otimes \dots \otimes \mathcal{H}_N$, where the blocks are simulated either in parallel on separate quantum devices or sequentially on the same machine.
While the qubits within one block can be highly entangled, we impose that the entanglement between blocks is weak, such that it can be recovered efficiently using classical resources.

\subsection{Projected variational quantum dynamics}
We perform the dynamics of the system governed by a Hamiltonian $H$ using the projected variational quantum dynamics (PVQD) algorithm~\cite{Barison_2021}. While traditional trotterized time evolution requires circuits that grow in depth with increasing evolution time $t$, the advantage of variational algorithms such as PVQD is that the circuit depth remains constant over the whole evolution. PVQD evolves the parameters $\theta$ of a quantum circuit ansatz $\ket{\psi(\theta)}$ in time, by minimizing the infidelity
\begin{equation}
\label{eq:infidelity_min}
    \theta_t = \argmin_\theta\left[ 1 - \big|\bra{\psi(\theta)}e^{-i\Delta t H}\ket{\psi(\theta_{t-1})}\big|^2\right]
\end{equation}
at every time step $t$. This ensures that $\ket{\psi(\theta_t)}$ is the state within the manifold defined by the ansatz that is closest to the true time-evolved state $e^{-i\Delta t H}\ket{\psi(\theta_{t-1})}$. Here, the time evolution unitary $e^{-i\Delta t H}$ can be expanded into gates using the Trotter-Suzuki decomposition of the first order for which the introduced error scales as $\cO(\Delta t^2)$. In our case, the time step $\Delta t$ is chosen to be small to keep the error negligible.

Crucially, PVQD only requires measuring fidelities between two quantum states. This can be achieved by sampling from hardware efficient circuits, in contrast to other variational methods such as the time-dependent variational principle (TDVP)~\cite{dirac_1930,frenkel_1934, mclachlan_1964, Yuan2019theoryofvariational}, where complex-valued state overlaps need to be measured using for example Hadamard-tests.

The fidelity between two quantum circuits is usually obtained using the compute-uncompute method~\cite{Havlicek_2019}, in which one measures the probability of retrieving the all-zero bit string after evolving the circuit in~\Cref{eq:infidelity_min}. The optimization of this global loss function is known to be prone to cost function-dependent barren plateaus~\cite{Cerezo_2021}, i.e. the gradients vanish exponentially fast in the number of qubits $n$. It has been shown that for small enough time steps $\Delta t$, PVQD is not affected by this problem as the initial guess $\ket{\psi(\theta_{t-1})}$ has a non-zero overlap with the target state $e^{-i\Delta t H}\ket{\psi(\theta_{t-1})}$~\cite{haug2021optimal, Barison_2021}. In addition, in the following experiments, we further increase the variance of the gradient by measuring a local observable with the same maximum as the global fidelity. The observable is defined as averaging over the local $\ket{0}\bra{0}$ projectors
\begin{equation}
    \label{eq:local_fid}
    \mathcal{O}_{\text{loc}} = \frac{1}{n}\sum_{k=1}^n \mathbbm{1}^{\otimes k-1} \otimes \ket{0}\bra{0} \otimes \mathbbm{1}^{\otimes n-k}.
\end{equation}

\subsection{Circuit knitting}

Performing any measurements on the variational state defined on the composite Hilbert space $\mathcal{H} = \mathcal{H}_1 \otimes \mathcal{H}_2 \otimes \dots \otimes \mathcal{H}_N$ requires running circuits spanning across all blocks. To realize measurements on circuits of smaller sizes, we utilize circuit knitting techniques~\cite{Bravyi_2016,Peng2020, Mitarai_2021, Mitarai2021overheadsimulating, piveteau2023circuit} to cut cross-block gates and recover the entanglement using additional circuit evaluations and classical post-processing. Circuit knitting allows decomposing a global quantum channel $\mathcal{U}$ acting on a quantum state $\rho$ into locally realizable quantum channels $\mathcal{E}_k^i$ according to a quasi-probability decomposition (QPD)
\begin{equation}
    \mathcal{U}[\rho] = \sum_{k=1}^K \alpha_k \mathcal{E}_k^1 \otimes \mathcal{E}_k^2 \otimes \dots \otimes \mathcal{E}_k^N[\rho],
\end{equation}
for $K\in\mathbb{N}$ and $\alpha_k \in \R$.
In our specific case, $\mathcal{U}$ will be the channel defined by a unitary gate acting on qubits of separate blocks $\mathcal{H}_i \otimes \mathcal{H}_j$, $\rho = \ket{\psi}\bra{\psi}$ is the pure state defined by the circuit prior to applying this gate, and $\{\mathcal{E}_k^i, \mathcal{E}_k^j\}$ are the corresponding set of channels that act locally only within each subsystem $\mathcal{H}_i$ or $\mathcal{H}_j$.

In practice, for every circuit evaluation, the global channel $\mathcal{U}$ is replaced by some locally realizable channel $\mathcal{E}_k = \mathcal{E}_k^1 \otimes \dots \otimes \mathcal{E}_k^N$ sampled according to the probability distribution defined by $p_k \propto |\alpha_k|$. While the QPD provides an unbiased estimator of the true expectation value of the measurement, the sampling cost required to achieve the same precision increases. Crucially, some of the $\alpha_k$ can be negative, which leads to a sampling overhead of 
\begin{equation}
    \omega(\mathcal{U}, \{\mathcal{E}_k^i\}_{k,i}) = \left(\sum_k |\alpha_k| \right)^2.
\end{equation}
This overhead is multiplicative\footnote{In general, the overhead is sub-multiplicative as, for the combination of multiple gates, a more efficient QPD can be found~\cite{piveteau2023circuit,schmitt2023cutting,ufrecht2023optimal}. In order to allow for a straightforward implementation of the circuit knitting scheme, in the following, we nevertheless assume a multiplicative sampling overhead.} and, hence, scales exponentially in the number of gates that are cut. 

\subsection{Overhead constrained PVQD}
\begin{figure}
    \centering
    \includegraphics{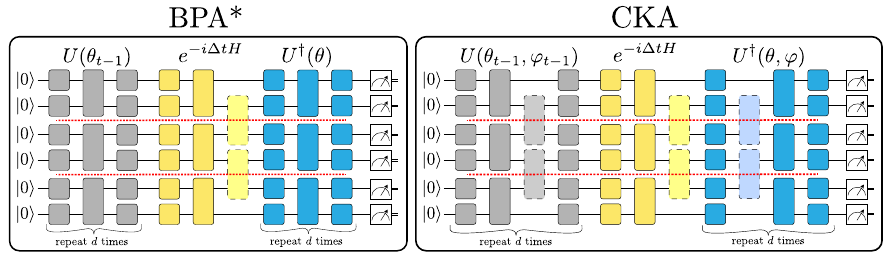}
    \caption{Circuits used to measure the (local) fidelity between the time evolved state $e^{-i\Delta t H} U(\theta_{t-1}, \phi_{t-1}) \ket{0} = e^{-i\Delta t H}\ket{\psi(\theta_{t-1},\phi_{t-1})}$ and the ansatz $\ket{\psi(\theta, \phi)}$ required for a PVQD optimization step. We cut the circuits into distinct blocks (indicated by the red dashed lines), which can each be simulated on a separate quantum device. In the experiments considered in this work, the single-qubit gates are realized using $R_X$ rotations, while the two-qubit gates correspond to $R_{ZZ}$ rotations. Gray-shaded gates are fixed by the parameters of the last time step $t$, while the parameters of the blue-shaded gates are varied to optimize~\Cref{eq:min_cka}. This structure is repeated $d$ times to increase the expressibility of the ansatz.The trotterized time evolution unitaries are colored yellow. \textbf{Left panel:}~Block product ansatz (BPA*) where the only entangling gates between different blocks appear in the Trotter step. (In a pure block product approximation (BPA) all inter-block gates are omitted, including the ones arising in the time evolution step.) \textbf{Right panel:}~Circuit knitting ansatz (CKA). Here, additional entangling gates between the different blocks are introduced into the ansatz. For clarity, the parameters of these dashed, light-colored gates are labeled $\phi$, whereas $\theta$ denotes the angles of all other gates that do need to be cut.}
    \label{fig:ansatz}
\end{figure}

The circuit that needs to be run to evaluate the fidelity in~\Cref{eq:infidelity_min} is composed of gates arising from the Trotter step unitary $e^{-i\Delta t H}$ and gates in the variational ansatz state $\ket{\psi(\theta)}= U(\theta)\ket{0}$ that potentially span across multiple blocks and thus have to be cut (see Fig.~\ref{fig:ansatz}). 
For the Trotter gates, we restrict the analysis to 2-local Hamiltonians, such that the multiqubit gates appearing in the Trotter expansion are given by two-qubit rotations defined as $e^{-i\Delta t J_{ij} \sigma_i \otimes \sigma_j}$, for Pauli operators $\sigma_i, \sigma_j\!\in\!\{X,Y,Z\}$ and coupling coefficients $J_{ij} \in \R$\footnote{This case includes widely studied spin-1/2 Hamiltonians like the Ising or Heisenberg models. However, our framework can be extended to Hamiltonians with k-local interactions.}.  The sampling overhead imposed by cutting a single instance of this gate with the optimal decomposition is given as $\omega_{J_{ij}} = \big(1 + 2|\sin(2\Delta t J_{ij})|\big)^{2}$~\cite{Mitarai2021overheadsimulating, piveteau2023circuit}. For the time evolution to be accurate, we require $\Delta t$ to be small. Moreover, we consider only cases in which the coupling $J_{ij}$ between qubits of different blocks is weak. Hence, we can assume $\Delta t J_{ij} \ll 1$, and thus, $\omega_{J_{ij}}$ is close to 1. If the Trotter step requires a total of $L$ such gates to be cut, the overhead scales as $\omega_{\Delta t} = \omega_{J}^{L}$, where for simplicity, we take $J_{ij} = J\ \forall ij$. While this scales exponentially in the number of gates, the base is small, and for a finite number of blocks, the overhead remains manageable.

For the cross-block gates introduced by the variational state $U(\theta)$, the analysis is less straightforward, as generally, the ansatz can be constructed from an arbitrary gate set. 
Many commonly used ansatzes consist of parameterized single-qubit rotations followed by CNOT gates that impact the entanglement. Cutting a CNOT gate, however, comes at a fixed cost of $\omega_{\text{CNOT}} = 9$. Even when employing more intricate cutting schemes that reduce the overhead of cutting $n$ CNOT gates simultaneously, the sampling overhead grows as $\omega_{\text{CNOT}^{\otimes n}} = (2^{n+1}-1)^2$~\cite{schmitt2023cutting}.
An alternative class of two-qubit gates that allow more control over the sampling overhead when being cut are parameterized two-qubit rotations such as those appearing in the Trotter decomposition. If  one cuts $M$ two-qubit rotations with angles $\phi_1,\dots\phi_M$, the multiplicative sampling overhead needed to evaluate the PVQD loss function with the circuit knitting scheme is given as
\begin{equation}
    \label{eq:tot_overhead}
    \omega(\phi) = \omega_{\Delta t}\cdot \left(\prod_{i=1}^M\big(1 + 2|\sin(\phi_i)|\big)^{2}\right)^2,
\end{equation}
where $\omega_{\Delta t}$ is the overhead due to cutting the Trotter step, and the additional square appears due to doubling the circuit (see~\Cref{fig:ansatz}). The total overhead can become extremely large if the angles $\phi_i$ are unbound. A way to circumvent this issue is to employ a block product ansatz that does not introduce any entangling gates between different blocks. This ansatz is shown in~\Cref{fig:ansatz} on the left and labeled as BPA* to distinguish it from a pure block product approximation (BPA) where also the entangling Trotter gates are omitted. While the BPA* comes at a minimal sampling overhead, it is not able to capture any entanglement between different blocks. Even for weakly entangled systems, the ansatz is thus expected to fail after evolving the system for a long enough time. Hence, it becomes necessary to add parameterized entangling gates between different blocks of the ansatz state (see right panel of \Cref{fig:ansatz}). We refer to this type of ansatz as circuit knitting approximation (CKA).

\begin{figure}
    \centering
    \includegraphics[width=.75\columnwidth]{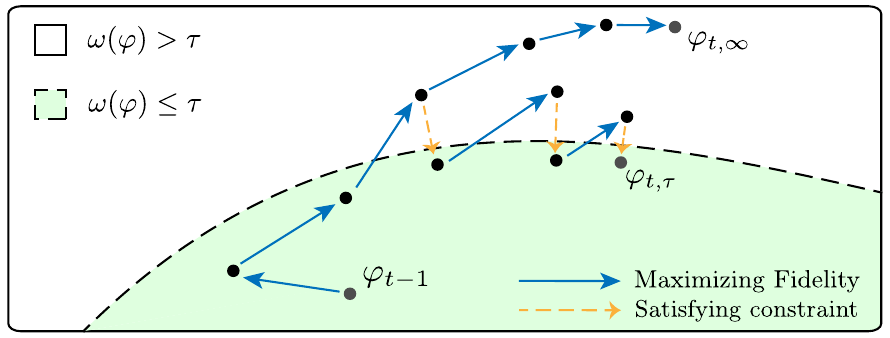}
    \caption{Solving the constrained optimization problem defined in~\Cref{eq:min_cka} to evolve the ansatz state by one-time increment. The optimization starts at the parameters of the last time step $t-1$. In every iteration, the parameters are first updated to maximize the fidelity with respect to the true time evolved state using an ADAM~\cite{kingma2017adam} update step (blue arrow). After the update, the multiplicative sampling overhead $\omega(\varphi)$ is computed according to~\Cref{eq:tot_overhead} and compared against the threshold $\tau$. In case $\omega(\varphi) > \tau$, the parameters are projected onto the manifold of $\omega(\varphi) \leq \tau$ (orange dashed arrow). This procedure is repeated until the parameters converge; the final point is labeled as $\varphi_{t,\tau}$. In contrast, the path on the top represents the usual, unconstrained optimization with no predefined threshold that converges to different parameters $\varphi_{t,\infty}$ which, however, incur an uncontrolled sampling overhead.}
    \label{fig:sketch}
\end{figure}

In order to keep the overhead $\omega$ controllable throughout the optimization of the CKA, we add a constraint to the optimization of~\Cref{eq:infidelity_min} such that $\omega$ is always bound by a threshold~$\tau > 1$
\begin{align} 
\label{eq:min_cka}
\theta_t, \varphi_t =& \argmin_{\theta,\phi}\left[ 1 - \big|\bra{\psi(\theta, \varphi)}e^{-i\Delta t H}\ket{\psi(\theta_{t-1}, \varphi_{t-1})}\big|^2\right] \\
\textnormal{s.t.} \, & \omega(\varphi) \leq \tau \nonumber,
\end{align}
where we denote by $\theta$ the parameters of gates acting within a single block and by $\phi$ the parameters of gates that are being cut, i.e. two-qubit gates stretching across two blocks. 
We satisfy the constraint throughout the optimization by projecting the parameters $\phi$ back into the allowed subspace defined by $\omega(\varphi) \leq \tau$ (see~\Cref{fig:sketch}). This projection is performed after every PVQD update step, which would result in circuits exceeding the predefined overhead threshold. We note that the number of entangling gates is fixed by the ansatz structure and the overhead is controlled by tuning the values of the parameter in those gates. In some cases, this might lead to gates being effectively removed from the circuit when the rotation angles are set to 0 during the optimization. This behavior is analyzed in detail in~\Cref{app:params}. The steps of the algorithm are outlined in Algorithm~\ref{alg:constr_opt}, and an in-depth description is provided in~\Cref{app:opt}.

\section{Results}
\label{sec:results}
As an example application of our method, we consider the transverse field Ising model (TFIM) spin system
\begin{equation}
    H = \sum_{\langle ij \rangle}J_{ij}Z_iZ_j +  \sum_{i} X_i,
\end{equation}
where we assume that the coupling between neighboring spins $J_{ij}$ is large for $i,j$ in the same block and small for $i, j$ in different blocks. In all subsequent simulations, we start the time evolution from the product state $|0\rangle^{\otimes n}$ of all $n$ spins pointing up. We compare our circuit knitting ansatz (CKA) with different thresholds $\tau$ to a pure block product approximation (BPA) and to the BPA*, where the full Trotter step, including all cross-block interactions, is implemented.

\begin{figure}
    \centering
    \includegraphics[width=0.75\columnwidth]{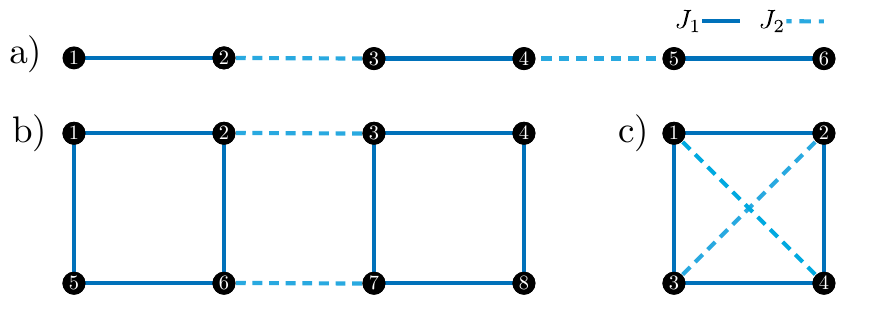}
    \caption{Spins in a transverse field Ising model. In (a) and (b), the system is split into blocks, where the coupling between two blocks $J_2$ is much smaller than the coupling within a block $J_1$. In (c), the strong coupling corresponds to nearest-neighbor interactions, whereas the next-nearest neighbor interactions are weak.}
    \label{fig:setup}
\end{figure}

\subsection{Spin chain}
\label{sec:ising_1d}
In the first experiment, we consider a spin chain of $N=3$ blocks of 2 spins each, as shown in~\Cref{fig:setup}~(a). The coupling within one block is chosen to be at the critical point $J_{ij} = J_1 = 1$, whereas the coupling between two blocks is set to $J_{ij}=J_2 = 1/4$. The ansatz follows the structure of the Trotter decomposition of $e^{-i\Delta tH}$ with $d=3$ repetitions of alternating layers of $R_X$ and $R_{ZZ}$ rotations (see~\Cref{fig:ansatz}). The total number of $R_{ZZ}$ gates that need to be cut is $N - 1$ for the BPA* and $(2d + 1)(N-1)$ for the CKA.

\begin{figure}
    \centering
    \includegraphics[width=\columnwidth]{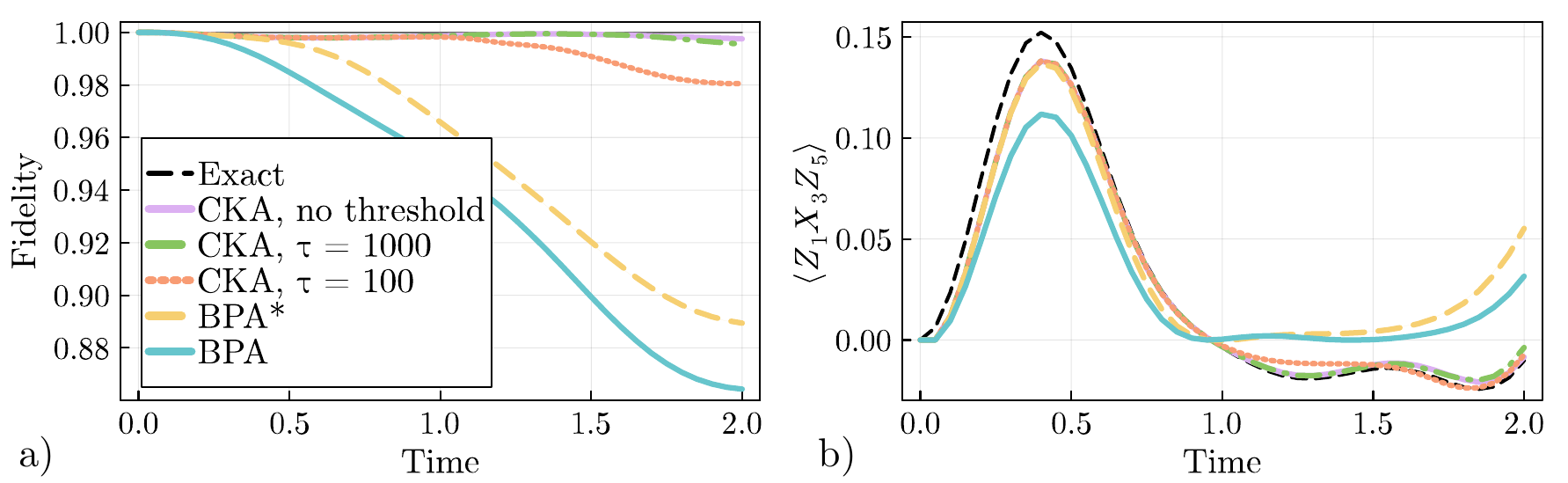}
    \caption{
    Simulating the dynamics of a TFIM spin chain consisting of 3 blocks with 2 spins each. \textbf{(a)} Fidelity of our time evolved ansatz with respect to the exact solution. \textbf{(b)} Expectation value of the observable acting as $Z$ on spins 1 and 5 and as $X$ on spin 3. We compare an ansatz involving parameterized two-qubit gates between blocks that are cut (CKA) while keeping the sampling overhead controlled under a threshold $\tau$ and a block-product state ansatz without entangling gates between the blocks (BPA). In the case of the latter, we further differentiate between optimizing with a block-product Trotter gate (i.e, no inter-block interactions) or the full Trotter gate, including the exact inter-block interactions (BPA*). We find that CKA can reach higher fidelities at longer times while the exact accuracy can be controlled by changing the overhead threshold.}
    \label{fig:1d_ising}
\end{figure}

In~\Cref{fig:1d_ising}~(a) we plot the fidelity of time-evolved states obtained through PVQD state vector simulations with respect to the exact solution. We observe that the pure block product ansatz optimized with block product Trotter gates (BPA) has the poorest performance as the fidelity quickly drops and reaches a value of only 0.87 at time $t=2$. This behavior is, however, expected since neither the ansatz nor the optimization takes into account any interactions or entanglement between different blocks of the systems. Adding (and cutting) the Trotter gates involving cross-block interactions while keeping the same block product ansatz (BPA*) slightly increases the fidelity. Finally, we expand the ansatz itself by adding parameterized gates between the blocks which are cut (CKA), and employ the overhead-constrained PVQD algorithm for the evolution. We are able to control the fidelity by tuning the threshold hyper-parameter $\tau$ that constrains the allowed sampling overhead for the ansatz. Ultimately, our optimization scheme gives us the means to naturally interpolate between the results obtained with a block product ansatz which incurs only a minimal sampling overhead, and the unconstrained PVQD evolved state, which gives rise to an unbounded overhead.

In~\Cref{fig:1d_ising}~(b), we show the evolution of a correlated observable acting on all three blocks. The behavior of long-ranged observables is typically more difficult to capture in hardware-efficient variational simulations, as their support grows faster compared to purely local observables. The BPA(*) is expected to fail in representing correlations spanning across different blocks, as becomes evident at times $t>1$. In contrast, the CKA with the particular thresholds chosen here can capture the inter-block correlations accurately also for long times.

\begin{figure}[t]
    \centering
    \includegraphics[width=\columnwidth]{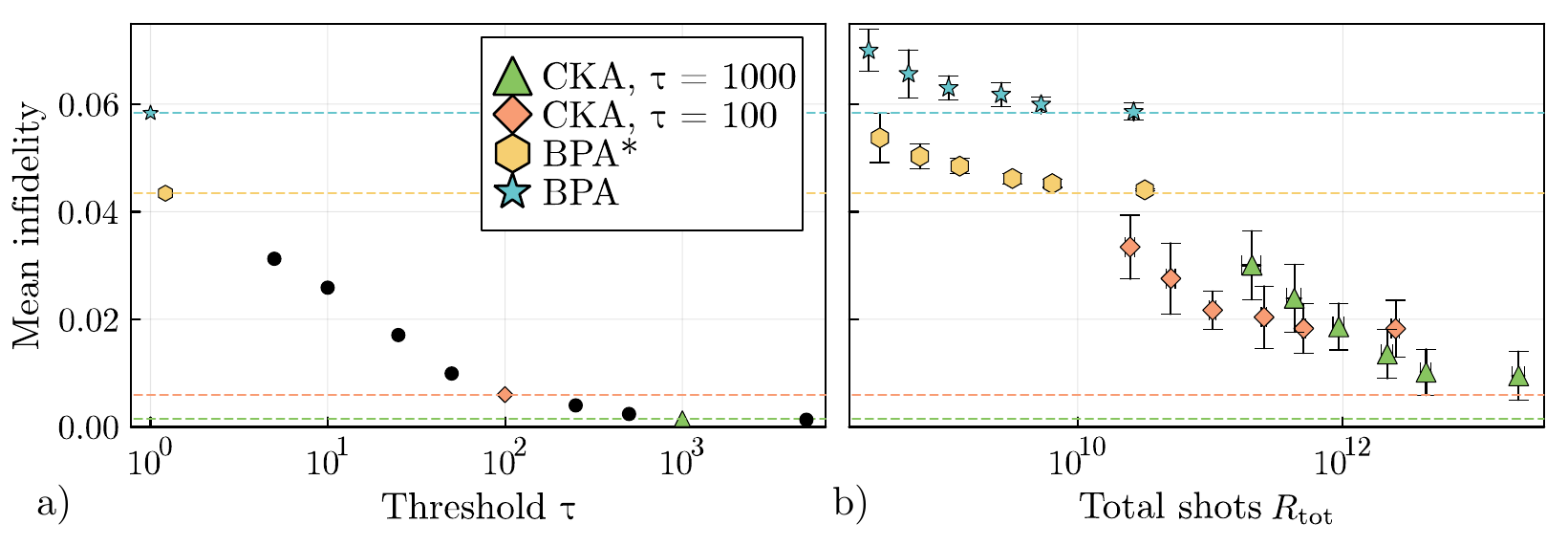}
    \caption{Mean infidelity over the time evolution. In \textbf{(a)} as a function of the threshold $\tau$ constraining the sampling overhead in the statevector simulations presented in~\Cref{fig:1d_ising}. The black points correspond to additional simulations performed (for thresholds 5, 10, 25, 50, 250, 500, 5000) to interpolate between the points shown in the other plots. In \textbf{(b)} as a function of the total shots required for the simulation, with dashed lines indicating the result achieved by the statevector simulation. Note that here we do not plot the CKA without a threshold, as the first point of this method would start at $10^{17}$ total shots and is, thus, unfeasible in reality.}
    \label{fig:shots}
\end{figure}

To explicitly see how the fidelity obtained with the overhead-constrained optimization increases with a higher threshold, we plot the mean infidelity
\begin{equation}
I = \frac{1}{T}\sum_{t=1}^T\left[1 - |\braket{\psi(\theta_t,\phi_t)}{\psi_t}|^2 \right]
\end{equation}
of the simulations with respect to the exact solution $\ket{\psi_t}$ in~\Cref{fig:shots}~(a). A larger overhead threshold improves the expressibility of the ansatz as the inter-block gate parameters are less constrained. As a result, the mean infidelity decreases. In order to fully quantify the computational cost required to achieve a certain fidelity, we further include a shot-based simulation, taking into consideration finite sampling noise. \Cref{fig:shots}~(b) shows how the mean infidelity decreases as the total number of shots is increased. For every point, 10 simulations were performed with a fixed number of shots $R$ per circuit evaluation. The total number of shots for every run is calculated as
\begin{equation}
    \label{eq:total_shots}
    R_{\text{tot}} = R\cdot n_{\text{iter}}\cdot 2\,n_{\text{params}} \cdot \sum_{t=1}^{T}\,\omega(\phi_t),
\end{equation}
where $n_{\text{iter}}=200$ is the number of iterations per time step\footnote{This number has been chosen high to ensure the optimization converges. Exploiting smart termination criteria should allow to significantly reduce the number of iterations.}, $2\cdot n_{\text{params}}$ the cost of calculating the gradient using the parameter shift rule for $n_{\text{params}}$ parameters, and $T=40$ the number of time steps in the simulation. The overhead is set to 1 for the BPA. While~\Cref{fig:shots}~(a) suggests that increasing the threshold improves the expressibility of the ansatz, leading to decreasing infidelities,~\Cref{fig:shots}~(b) demonstrates that shot noise limits the simulation from reaching the ideal infidelity. 
Given a fixed budget of total shots $R_{\text{tot}}$, choosing the optimal threshold $\tau$ and the number of shots $R$ per circuit evaluation is a nontrivial constrained optimization problem.
In~\Cref{fig:shots}~(a), this balance is illustrated as there is a regime around $10^{11}$ total shots where having a lower threshold (but larger $R$) results in lower infidelity than a high threshold (but smaller $R$). On the other hand, for a higher budget of around $10^{12}$ total shots, choosing the larger threshold is advantageous.

\begin{figure}
    \centering
    \includegraphics[width=\columnwidth]{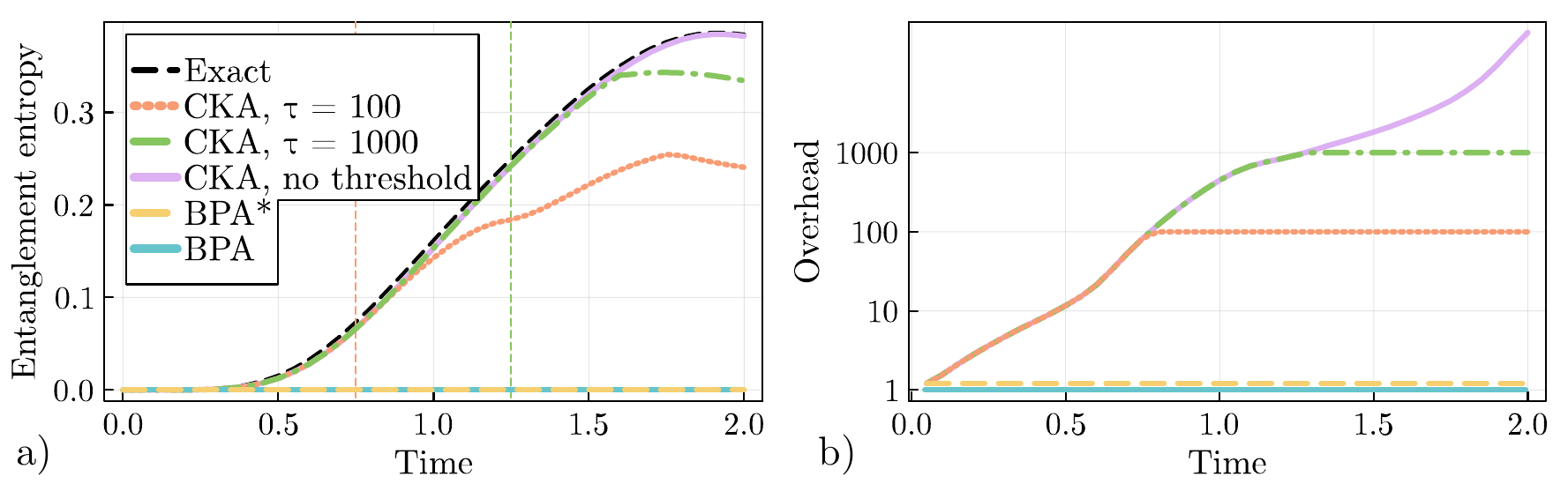}
    \caption{\textbf{(a)} The entanglement entropy between the center block and the outer two blocks is calculated as a function of time for different ansatzes. The overhead threshold in the CKA determines the time window for which the entanglement growth can be captured accurately. In contrast, the BPA(*) is not able to account for any inter-block entanglement by construction. \textbf{(b)} Required sampling overhead versus simulation time for different thresholds. We indicate the exact time at which the overhead reaches the threshold as vertical dashed lines in (a).}
    \label{fig:entanglement}
\end{figure}

We further investigate how entanglement between different blocks can be captured with the CKA and how the entanglement correlates to the sampling overhead. We generally expect that imposing a threshold on the sampling overhead required for the circuit knitting scheme limits the entanglement that can arise between the subsystems. In order to quantify the entanglement in our ansatz state, we split the system shown in~\Cref{fig:setup}~(a) into a bipartite system. We call $A$ the subsystem containing the center block and $B$ the subsystem containing the outer two blocks. We write the pure state $\ket{\psi}=U(\theta,\phi)\ket{0}$ defined by the quantum circuit in its Schmidt decomposition
\begin{equation}
    \label{eq:schmidt}
    \ket{\psi} = \sum_{k=1}^{\text{dim}(A)}\lambda_k\ket{a_k}\ket{b_k},
\end{equation}
where $\lambda_k \geq 0$ are the Schmidt coefficients, $\ket{a_k},\ket{b_k}$ the Schmidt basis states in systems $A$ and $B$, respectively. From this decomposition, the von Neumann entanglement entropy can be easily computed as~\cite{nielsen_chuang_2010}
\begin{equation}
    \label{eq:entropy}
    E(\ket{\psi}) = - \sum_{k=1}^{\text{dim}(A)} \lambda_k^2 \log(\lambda_k^2).
\end{equation}
In~\Cref{fig:entanglement}~(a), we show how the entanglement entropy grows in time for different ansatzes. As expected, the BPA(*) captures no entanglement between the distinct blocks, while the CKA without a threshold recovers the full entanglement of the exact solution. For the CKA with $\tau=100,1000$, we observe that the entanglement entropy eventually starts deviating and stays below its exact value as expected. To understand whether the errors in the entanglement entropy arise due to the constrained optimization problem, we also show how the sampling overhead increases over time and, if applicable, caps at the threshold (see~\Cref{fig:entanglement}~(b)). Interestingly, the entanglement entropy is growing even after the sampling overhead saturates (indicated by the vertical lines) and does not plateau to a specific value. In this case, the optimization learns that due to the multiplicative overhead it is more efficient to have few entangling gates with large angles compared to many gates with small angles. Once the threshold is reached, the entanglement generation thus starts to be concentrated on a few gates which allows the entanglement entropy to grow further. A detailed explanation for this behavior is provided in~\Cref{app:params}.

\subsection{Two-leg ladder}
Next, we demonstrate that our scheme can also be applied to lattice geometries beyond the simple 1d spin chain. To that end, we consider the Ising model on a two-dimensional extension of the chain as shown in~\Cref{fig:setup}~(b). Each of the two blocks is comprised of 4 spins and coupled to the other block via weak nearest-neighbor interactions. To simulate the dynamics of this system, we choose an ansatz layout reflecting the corresponding trotterized time evolution operator. Specifically, we use alternating layers of single-qubit $R_X$ rotations and $R_{ZZ}$ rotations, repeated $d=5$ times. The total number of $R_{ZZ}$ that need to be cut is $2$ for the BPA* and $2(2d + 1)$ for the CKA.

\begin{figure}
    \centering
    \includegraphics[width=\columnwidth]{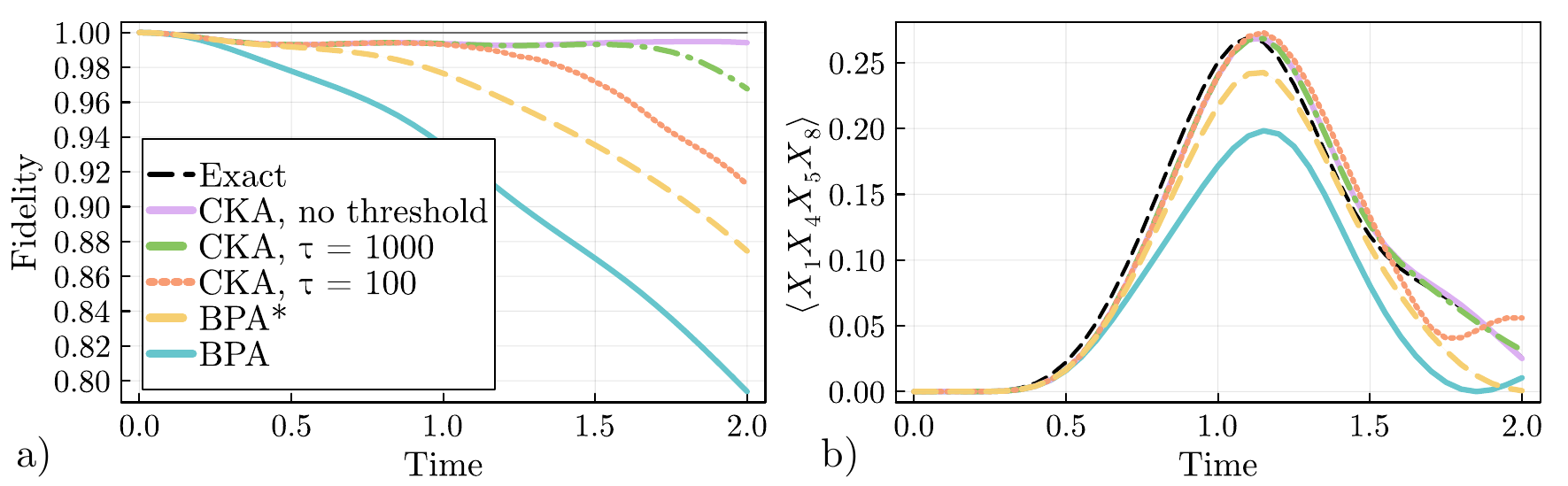}
    \caption{Simulating the dynamics of the two-leg ladder TFIM. \textbf{(a)} Fidelity of the simulation with respect to the exact solution for the BPA, BPA* and CKA with different thresholds. \textbf{(b)} Expected value of the observable acting as $X$ on the four outer qubits (cf.~in~\Cref{fig:setup}~(b)). The accuracy of the simulation improves as the threshold is increased.}
    \label{fig:two-leg-plot}
\end{figure}

In~\Cref{fig:two-leg-plot}, we show how the fidelity of the different ansatzes with respect to the exact solution evolves in time. Additionally, we plot the expectation value of the observable that acts as $X$ on the four outer qubits (the two qubits on the left of the first block and the two qubits on the right of the second block). While the block product ansatz (BPA*) initially tracks the qualitative behavior of the dynamics, it fails in the second half of the simulation period. Here, adding the cross-block entangling unitaries to the ansatz is necessary to accurately approximate the time-evolved state. The CKA with a threshold of $\tau = 100$ captures the qualitative dynamics of the observable plotted in~\Cref{fig:two-leg-plot} until $t\approx 1.5$. In order to accurately simulate the dynamics until $t=2$, the threshold has to be increased to $\tau = 1000$.

\subsection{Reducing circuit depth}
Many state-of-the-art quantum computing platforms such as those based on superconducting qubits feature only a limited qubit connectivity. Gates acting on qubits that are not adjacent in the device layout have to be implemented via additional two-qubit SWAP operations. However, these extra gates increase the amount of noise in a computation. In the era of NISQ devices, it is therefore crucial to find ways of reducing the circuit depth while keeping the simulations as accurate as possible. To that end, circuit knitting can be employed to cut long-range acting gates.

Here, we demonstrate the use of circuit knitting to effectively reduce the circuit depth in the variational simulation of the dynamics of the J1-J2 transverse field Ising model depicted in~\Cref{fig:setup}~(c) and defined by
\begin{equation}
    H = J_1\sum_{\langle ij \rangle}Z_iZ_j + J_2\sum_{\langle\langle ij \rangle\rangle}Z_iZ_j +  \sum_{i} X_i,
\end{equation}
where we again choose $J_1 =1$, $J_2 = 1/4$. $\langle ij \rangle$ indicates nearest-neighbors, whereas $\langle\langle ij \rangle\rangle$ corresponds to next-nearest-neighbors. Instead of cutting the system into blocks, we here cut the long-range gates induced by the next-nearest-neighbor interactions. 
We compare the PVQD dynamics for similar ansatzes as in the previous experiments with $d=4$ repeated layers. 
Specifically, we consider an ansatz that is composed only of hardware-efficient gates, i.e, gates acting only on nearest-neighbor spins/qubits. For consistency, we refer to this ansatz as BPA(*), even though we are not cutting the system into blocks in this case. In contrast, the CKA ansatz reflects the full interaction graph of the model and contains additional $4(2d +1)$ long-range entangling gates that are cut using circuit knitting. 

\begin{figure}
    \centering
    \includegraphics[width=\columnwidth]{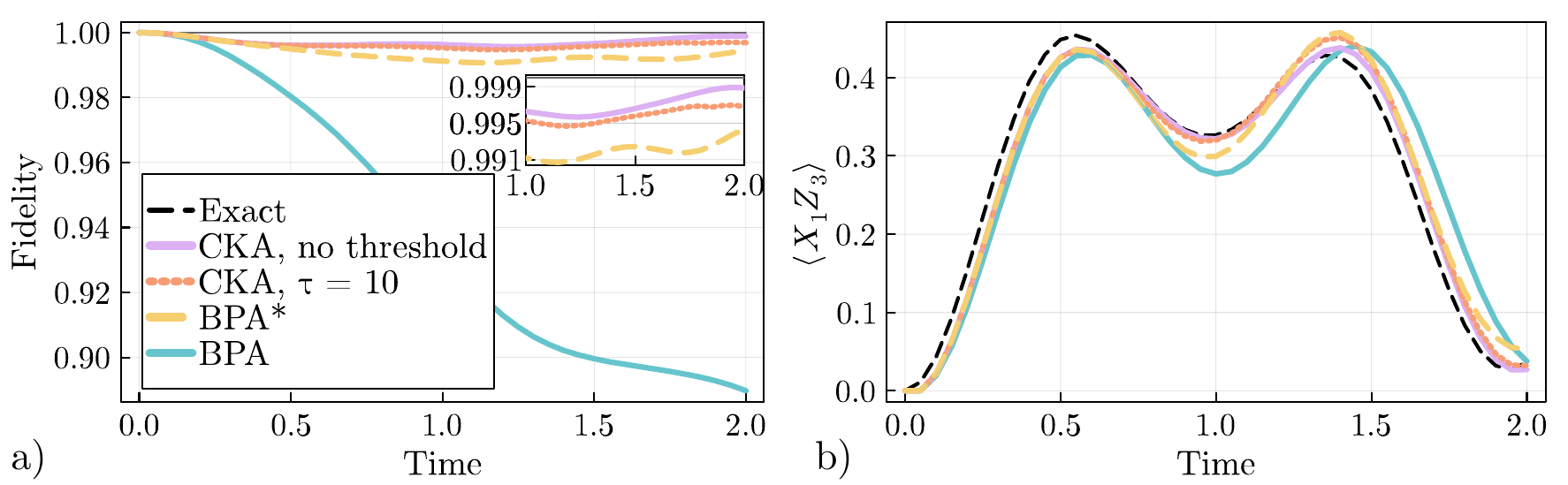}
    \caption{Simulating the dynamics of J1-J2 Ising model sketched in~\Cref{fig:setup} (c). We show \textbf{(a)} the fidelity of the simulation with respect to the exact solution as well as \textbf{(b)} the expectation value of the observable acting as $X$ on the upper left qubit and as $Z$ on the lower right qubit. BPA here indicates a hardware efficient circuit, whereas CKA includes next-nearest-neighbor 2-qubit gates that are cut using circuit knitting while the overhead is constrained by a threshold $\tau$. The main improvement over the BPA is given by BPA*, where next-nearest-neighbor interactions are considered in the Trotter decomposition of $e^{-i\Delta t H}$.}
    \label{fig:j1j2_plot}
\end{figure}

The results of our simulations are provided in~\Cref{fig:j1j2_plot}, where we show both the time dependence of the fidelity to the exact state and of an observable acting on two non-adjacent spins. In the BPA, all gates acting on next-nearest-neighbors are omitted from the circuit, including the Trotter step. As a result, the fidelity quickly deteriorates as we effectively evolve with a slightly different model where $J_2\!=\!0$. In contrast, for BPA*, the finite next-nearest-neighbor interactions are included in the Trotter step while the ansatz is kept hardware-efficient. In this case, the fidelities stay high throughout the time evolution interval. Hence, the hardware-efficient ansatz comprised of $d=4$ repeated layers is already able to accurately represent the long-range correlations and entanglement generated by the next-nearest-neighbor interactions of the model. However, we can improve on these fidelities even further by using the CKA with a comparatively small overhead threshold of $\tau=10$.

In order to quantify the depth reduction enabled by cutting long-ranged gates in this example, we count the number of SWAP gates required to run PVQD on this system without cutting any gates. Given a quantum device where the connectivity coincides with this geometry (i.e. nearest-neighbor spins/qubits are connected), every Trotter layer would require 2 SWAP gates. For the ansatz with 4 repeated layers, this results in $2\! +\! 2\!\cdot\! 4\!\cdot\! 2 = 18$ SWAP gates that can be saved using circuit knitting (2 for the Trotter step, $4\!\cdot\! 2$ for the ansatz, and the extra factor of 2 comes from doubling the circuit in the compute-uncompute method).

Overall, in this example, circuit knitting enables us to trade-off a larger circuit depth for an increased sampling overhead.

\section{Discussion \& Outlook}
\label{sec:conclusion}
Circuit knitting allows for the simulation of larger quantum systems using small quantum devices. While in general, cutting circuits can be expensive due to the sampling overhead, we show that this overhead can be controlled by constraining the parameters in the variational circuit optimization. Applying this technique to the dynamics simulation of spin systems with PVQD, we are able to achieve the optimal fidelity given a fixed budget of samples. A change in the threshold hyper-parameter $\tau$ leads to a trade-off between the accuracy of the simulation and the sampling overhead. The optimal threshold therefore depends on the quantum computing resources available, the desired accuracy, and the total evolution time.

In the examples considered in this work, we show that with a realistic sampling overhead, the accuracy of the dynamics simulations can be drastically improved compared to a simple block product ansatz. Classical resources can thus effectively be used to recover entanglement between the different subsystems. Our framework opens the door to simulating the dynamics of quantum systems with a large number of qubits that are otherwise not reachable with current hardware. Possible systems of interest are, for example, quantum impurities immersed in a bath~\cite{Kotliar_2006, Sun2016} or low-energy eigenstates of local lattice Hamiltonians and molecules~\cite{Eisert2010area,Schollw_ck_2011, liu2019vqe,McArdle2020}.
Furthermore, we show that our technique can also be used to reduce the circuit depth if, instead of cutting the system into blocks, we cut long-ranged but weak interactions. Here, we observed that with a controlled sampling overhead, the dynamics can be accurately simulated with hardware-efficient circuits.

Overhead-constrained circuit knitting allows for combining multiple quantum circuits that are individually already difficult to simulate classically, enabling the simulation of even larger systems given that the entanglement between subcircuits remains weak. This is of particular importance as the current trend in quantum hardware is to combine several smaller devices for distributed computing applications~\cite{ibm_roadmap}. Hence, our method presents another step towards the overarching goal of quantum utility~\cite{utility2023, ibm_zne2023}.

The direct application of circuit knitting to a trotterized simulation of dynamics, as performed in the recent utility experiments by IBM~\cite{utility2023} is prohibitively expensive. The reason is the accumulated sampling overhead when cutting a Trotter time evolution into subsystems. The sampling overhead is fixed by the coupling strength and scales exponentially with the simulation time and thus cannot be controlled to remain below a manageable threshold. This issue is addressed with our overhead-constrained circuit knitting approach applied to PVQD where we provide a way to control the total budget of circuit evaluations.

An expansion of this work would encompass a hardware experiment of the overhead-constrained PVQD. In this regard, it will be interesting to see whether current error mitigation techniques are powerful enough to mitigate the hardware noise to a level where the (local) fidelities can be measured to sufficient precision for the optimization to be successful.

Moreover, the constrained optimization presented in this work is not limited to PVQD but can be extended to arbitrary loss functions. As such, it could, for example, be applied to simulate ground states using circuit knitting and VQE~\cite{peruzzo2014, khare2023parallelizing} while keeping the sampling overhead controlled. More general, the overhead-constrained circuit knitting can be extended to any variational quantum algorithms where the total system can be split into weakly entangled blocks. In cases where the optimal partitioning of the system into subsystems cannot be physically motivated, heuristic methods might be applied to find the optimal placement of cuts~\cite{brandhofer2023optimal}. Overall, the question of where to optimally place the cuts is a non-trivial optimization problem on its own and represents an interesting direction for future research.

Finally, we remark that the calculations of the sampling overhead throughout this work are based on the worst-case scenario, where the total overhead is the product of the overheads required to cut individual gates.
It has recently been proposed that this overhead can be further reduced by using more intricate decompositions that cut multiple gates simultaneously~\cite{schmitt2023cutting, ufrecht2023optimal}. Alternatively, we could also take into consideration quantum communication between the different devices~\cite{Cuomo_2020}. In a recently appeared manuscript~\cite{gomez2023nearterm}, it has been shown that under similar conditions this can significantly increase the fidelity in distributed simulations of quantum dynamics. However, the hardware to implement a knitting scheme with quantum communication is currently missing and the additional computational costs of such a method will highly depend on how efficient and flexible these quantum links will be.

\paragraph{Code availability}
Simulations presented in this work were performed in Julia~\cite{Julia-2017} using the \texttt{Yao.jl} framework~\cite{Yao} and are available on Github~\cite{Code}.

\paragraph{Acknowledgments}
We thank Stefano Barison, Julien Gacon, and David Sutter for fruitful discussions on hybrid algorithms, optimization techniques, and circuit knitting. This research was supported by the NCCR MARVEL, a National Centre of Competence in Research, funded by the Swiss National Science Foundation (grant number 205602).

\bibliographystyle{quantum}
\bibliography{notes.bib}

\appendix

\section{Detailed description of the optimization}
\label{app:opt}
\begin{figure}
    \centering
    \includegraphics[width=0.7\columnwidth]{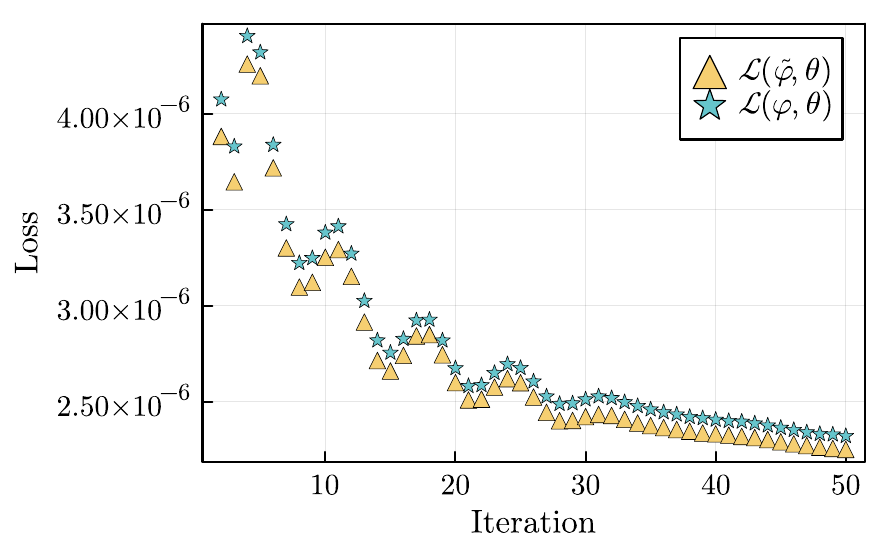}
    \caption{Example of a training curve given by the local infidelity loss function of~\Cref{eq:min_cka}, here for the 30th time step of the simulation with a threshold $\tau = 1000$ as shown in~\Cref{fig:1d_ising} of the main text. The blue stars indicate the final loss of each iteration, whereas the yellow triangles show the loss after the ADAM update but before projecting the parameters to the constraint-satisfying subspace. We only plot the first 50 of 200 iterations.}
    \label{fig:learning_curve}
\end{figure}

In this appendix, we provide a detailed description of the algorithm applied to optimize~\Cref{eq:min_cka}, including numerical values of hyper-parameters. An overview of the algorithm is given in the main text and in Algorithm~\ref{alg:constr_opt}. The initial guess for the new parameters for time step $t+1$ is given by $\theta^{k=0} = \theta_t + \Delta\theta$, where $\Delta\theta = \theta_t - \theta_{t-1}$ (the same procedure holds for the parameters $\phi$). For the first time step $t=1$, we set $\Delta\theta = 0$. Starting from this initial guess, we make an ADAM~\cite{kingma2017adam} update on $\theta^k, \phi^k$ according to the gradient of the objective function in~\Cref{eq:min_cka}, where the gradient is calculated using auto-differentiation for statevector simulations and using the parameter shift rule for shot based simulations. We further modify the ADAM algorithm slightly by keeping the momentum for the inter-block parameters $\phi$ at 0. This has been observed to speed up the convergence. Otherwise, we use standard hyper-parameters $\beta_1 = 0.9$ and $\beta_2 = 0.999$. The learning rate is set to $10^{-3}$ for statevector simulations and to $10^{-2}$ for shot-based simulations. This update step yields the next single-block parameters $\theta^{k+1}$ and a candidate for the cross-block parameters $\tilde{\varphi}^{k+1}$. 
The total sampling overhead is computed according to~\Cref{eq:tot_overhead} and compared against the threshold $\tau$. If $\omega(\tilde{\varphi}^{k+1}) \leq \tau$, the parameters are accepted and we continue with the next iteration. If, however, the overhead is larger than the threshold, the cross-block parameters $\phi$ are projected back to the region where $\omega(\phi) \leq \tau$ using the following procedure. The gradient $g = \nabla_\phi\omega(\tilde{\varphi}^{k+1})$ is calculated using auto-differentiation. Using a step size of $\mu = 10^{-3}$ (for the 1d statevector experiments we set $\mu = 10^{-5}$), we perform $m \in \N$ steps from $\tilde{\varphi}^{k+1}$ along $g$ until
\begin{equation}
    \omega(\tilde{\varphi}^{k+1} - m\cdot\mu\cdot\frac{g}{\norm{g}}) \leq \tau
\end{equation}
is fulfilled. The new cross-block parameters are then defined as $\phi^{k+1} =\tilde{\varphi}^{k+1} - m\cdot\mu\cdot\frac{g}{\norm{g}}$, ensuring that the next optimization step starts in the constraint satisfying region. This procedure is repeated until convergence; in our simulations we ran the optimizations for 200 iterations. The number of (purely classical) steps required to project the parameters back to the constraint satisfying region is in the order of $m \approx 1$ for $\mu = 10^{-3}$ and $m \approx 50$ for $\mu = 10^{-5}$. An example of a learning curve resulting from this optimization procedure is given in~\Cref{fig:learning_curve}.

\begin{algorithm}
\caption{Algorithm employed to solve the constrained optimization problem defined in~\Cref{eq:min_cka} in order to perform a time step of the overhead-constrained PVQD.}
\label{alg:constr_opt}
	\begin{algorithmic}[1]
        \STATE $\theta^0, \varphi^0 \gets \theta_{t-1}, \phi_{t-1}$
        \STATE $k \gets 0$
		\WHILE{algorithm not converged}
        \STATE $k \gets k + 1$
		\STATE $\theta^k, \tilde{\phi}^k \gets $ ADAM~\cite{kingma2017adam} update step on the objective function in~\Cref{eq:min_cka}
		\IF{ $\omega(\tilde{\varphi}_k) > \tau$}
		\STATE $g \gets \nabla_\varphi \omega(\tilde{\varphi}^k)$
        \STATE $\mu^* \gets \min \{ \mu > 0\, |\, \omega(\tilde{\varphi}^k - \mu g) \leq \tau \}$
        \STATE $\varphi^k \gets \tilde{\varphi}^k - \mu^* g$
		\ELSE
		\STATE $\varphi^k \gets \tilde{\varphi}^k$
		\ENDIF

		\ENDWHILE
        \STATE $\theta_{t}, \phi_{t}  \gets \theta^k, \varphi^k$
	\end{algorithmic}
\end{algorithm}

\section{Parameter behavior in overhead-constrained PVQD}
\label{app:params}
\begin{figure}
    \centering
    \includegraphics[width=\columnwidth]{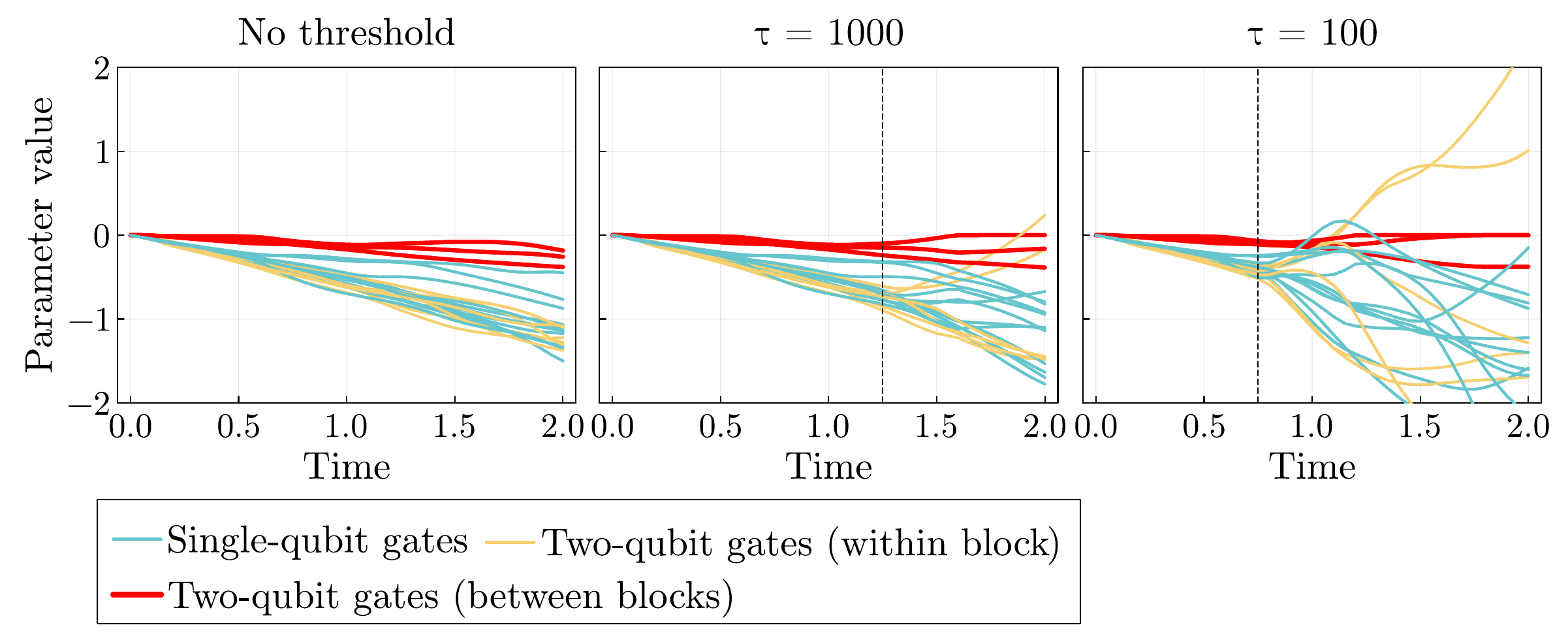}
    \caption{Parameter evolution during simulations of the Ising chain as shown in~\Cref{fig:1d_ising} of the main text for CKA runs without threshold and with thresholds $\tau = 100,\ 1000$. The vertical black dashed lines indicate the exact time when the sampling overhead reaches the imposed threshold. We differentiate between angles parameterizing the gates between blocks (referred to as $\phi$ in the main text) and parameters of the single-qubit and remaining two-qubit gates (referred to as $\theta$). For the CKA simulation without a threshold, the parameters evolve smoothly throughout the evolution. When a threshold is imposed, the parameter evolution becomes more involved once the threshold has been reached. Furthermore, to limit the (multiplicative) overhead, the algorithm effectively removes some of the inter-block gates by reducing their parameters to zero.}
    \label{fig:params}
\end{figure}

In this appendix, we analyze how the circuit parameters evolve during the overhead-constrained PVQD, explaining how the entanglement between subsystems is able to grow even after the threshold on the sampling overhead has been reached. \Cref{fig:params} shows the circuit parameters during the time evolution of the Ising chain discussed in~\Cref{sec:ising_1d}. The three CKA simulations are identical until the sampling overhead saturates at the threshold $\tau$ (indicated by vertical, dashed lines). At this point, the parameters $\phi$ of the gates between blocks are no longer freely optimized but are constrained such that the overhead $\omega(\phi)$ remains below the threshold. 

Nevertheless, \Cref{fig:entanglement} in the main text shows that the entanglement entropy increases after the threshold has been reached. This is achieved by reducing the angles in the entangling gates for two of the three layers of the ansatz, allowing the angles in the remaining layer to increase further. The optimization algorithm thus learns that concentrating the generation of entanglement onto one layer reduces the sampling overhead. This is due to the fact that the overhead is multiplicative (see~\Cref{eq:tot_overhead} in the main text).

Indeed, for $\tau = 100$ only one layer is parameterized by non-zero angles at the end of the evolution. For the $\tau = 1000$ case, a similar development is observed, albeit not as extreme (only for one layer the angles become zero). The single-qubit gates and two-qubit gates within a block are optimized to accommodate this transition, leading to a more intricate evolution of the parameters.

\end{document}

%% file: structure.tex
\usepackage[affil-it]{authblk}
\usepackage[usenames,dvipsnames]{xcolor}
\usepackage{amsmath,amsthm,amssymb,dsfont,pifont}
\usepackage{enumerate}
\usepackage[english]{babel}
\usepackage{graphicx}	
\usepackage{subcaption}
\usepackage[margin=3cm]{geometry}
\usepackage{url}
\usepackage{todonotes}
\usepackage{bbm}
\usepackage{caption}
\usepackage{subcaption}
\usepackage{tikz}
\usetikzlibrary{chains}
\usetikzlibrary{fit}
\usetikzlibrary{quantikz}
\usepackage{makecell}
\usepackage{cite}

\usepackage{epsfig}
\usetikzlibrary{shapes.symbols,patterns} 
\usepackage{pgfplots}

\usepackage{hyperref}[breaklinks]
\hypersetup{colorlinks=true,citecolor=blue,linkcolor=blue,filecolor=blue,urlcolor=blue}

\usepackage{nicefrac}
\usepackage{mathtools}

\usepackage{algorithm}
\usepackage{algorithmic}

\usepackage{optidef}

\usepackage{bold-extra}

\tikzset{meter/.append style={draw, inner sep=8, rectangle, font=\vphantom{A}, minimum width=30, line width=.8,
 path picture={\draw[black] ([shift={(.1,.3)}]path picture bounding box.south west) to[bend left=50] ([shift={(-.1,.3)}]path picture bounding box.south east);\draw[black,-latex] ([shift={(0,.1)}]path picture bounding box.south) -- ([shift={(.3,-.1)}]path picture bounding box.north);}}}
 
\theoremstyle{plain}

\theoremstyle{definition}

\newcommand*{\cO}{\mathcal{O}}

\newcommand*{\N}{\mathbb{N}}

\newcommand*{\R}{\mathbb{R}}

\renewcommand{\phi}{\varphi}

\newcommand{\norm}[1]{\left\lVert#1\right\rVert}

\definecolor{mylightgreen}{RGB}{219,255,192}
\definecolor{mylightblue}{RGB}{240,255,252}
\definecolor{mylightyellow}{RGB}{255,248,180}
\definecolor{mylightorange}{RGB}{252,248,236}
\definecolor{mygraywhite}{RGB}{243, 243,243}
\definecolor{mylightred}{RGB}{255, 209,192}
\definecolor{mylightpink}{RGB}{255,240,252}

\usepackage{float}

\usepackage{physics}
\usepackage{algorithmic}
\usepackage{algorithm}
\usepackage{cleveref}
\DeclareMathOperator*{\argmin}{arg\,min}

\usepackage{tikz}
\usetikzlibrary{quantikz}
\usetikzlibrary{shapes.symbols,patterns} 
\usepackage{pgfplots}
\usetikzlibrary{shapes.symbols,patterns} 

\definecolor{myBlue}{RGB}{31,119,180}
\definecolor{myOrange}{RGB}{255, 127, 14}
\definecolor{myGreen}{RGB}{44,160,44}
\definecolor{myRed}{RGB}{214,39,40}
\definecolor{myViolet}{RGB}{148,103,189}
\definecolor{myGrey}{RGB}{127,127,127}
\definecolor{myCyan}{RGB}{23, 190, 207}

\definecolor{featureMap}{RGB}{23, 190, 207}
\definecolor{variationalForm}{RGB}{148,103,189}
\definecolor{highlight}{RGB}{214,39,40}

\usepackage[parfill]{parskip} 

\allowdisplaybreaks